\documentclass[8pt,twoside]{article}  
\usepackage{times}
\usepackage{graphicx}
\usepackage[round,numbers,sort&compress]{natbib}

\newcommand{\degree}{\ensuremath{^\circ}}

\begin{document}

\twocolumn[{\Huge \textbf{Melting of Single Lipid Components in Binary Lipid Mixtures: A Comparison
  between FTIR Spectroscopy, DSC and Monte Carlo Simulations\\*[0.1cm]}}\\

{\LARGE{\center M. Fidorra$^{\dag, \ddag}$, T. Heimburg$^{\dag}$ and
    H.M. Seeger$^{\dag,\star,\ast}$\\*[-0.2cm]}
{\normalsize\emph{ \center 
$^{\dag}$The Niels Bohr Institute, University of
Copenhagen, Copenhagen, Denmark\\
$^{\ddag}$ MEMPHYS-Center of Biomembrane Physics, Department of
  Physics,\\
University
of Southern Denmark, Odense, Denmark\\
$^\star$ CNR-INFM National Research Center on 'nanoStructures and bioSystems at
  Surfaces S3',\\Modena, Italy\\*[0.3cm]}}


{\normalsize Monte Carlo (MC) Simulations, Differential Scanning Calorimetry (DSC) and Fourier Transform InfraRed (FTIR)
spectroscopy were used to study the melting behavior of single lipid
components in two-component membranes of
1,2-Dimyristoyl-D54-sn-Glycero-3-Phosphocholine (\textit{DMPC-d54}) and
1,2-Distearoyl-sn-Glycero-3-Phosphocholine (\textit{DSPC}). Microscopic
information on the temperature dependent melting of the 
single lipid species could be investigated using FTIR. The microscopic
behavior measured could be well described by the results from the MC
simulations. These simulations also
allowed to calculate heat capacity profiles as determined with DSC. These ones provide
macroscopic information about melting enthalpies and entropy changes which are
not accessible with FTIR. Therefore, the MC simulations allowed us to link the two different experimental approaches of FTIR and DSC. }\\*[0.3cm]}

\footnotesize{\textbf{keywords:} domain formation; phase separation; FTIR; MC
  simulation; DSC; melting of single lipid components\\*[0.3cm]}]

\footnotetext{\footnotesize{\textbf{Abbreviations:} DSC, differential scanning
calorimetry; FTIR, fourier transform infrared
  spectroscopy; MC simulation, Monte Carlo simulation; DMPC, 1,2-dimyristoyl-sn-glycero-3-phosphocholine; DMPC-d54,
  1,2-dimyristoyl-d54-sn-glycero-3-phosphocholine; DSPC, 1,2-distearoyl-sn-glycero-3-phosphocholine}}

\footnotetext{$^\ast$\footnotesize{ Corresponding author:
heiko.seeger@unimore.it, Via G. Campi 213/A, 41100 Modena, Italy}}


\section*{Introduction}

Biological membranes display a complex lipid composition. The reason for the
big variety in lipid species is still not clear. It has been shown for
many different bacteria that their lipid synthesis depends on physical parameters
such as growth temperature or hydrostatic pressure~\cite[regard also the
citations  given therein]{Marr:EffTempEColi, Johnston:BrainLipidTemp,
Kleinschmidt:TemplipcompCyanidium, Sinensky:tempcontrolPhosphoLipSyn,
Cronan:MetaEColi, jain:platemodel, Cronan:MolBiolBacMemLip,
Johnston:VeillonellaandMegasphaera, DeLong:AdapMemLipDeepSeaBact,
Bartlett:PressureEffectsMicrobialProc, Koenneke:growthtempsuolph} . It
could be shown that the membranes of bacteria grown at higher temperatures
contain more saturated fatty acid chains and chains tend to be
longer~\cite[see also citations therein]{Marr:EffTempEColi,
Johnston:BrainLipidTemp, Kleinschmidt:TemplipcompCyanidium,
Sinensky:tempcontrolPhosphoLipSyn, Cronan:MetaEColi, jain:platemodel,
Cronan:MolBiolBacMemLip,Johnston:VeillonellaandMegasphaera,
Koenneke:growthtempsuolph}. Higher hydrostatic pressure induces an increased
synthesis of unsaturated lipids~\cite{DeLong:AdapMemLipDeepSeaBact,
Bartlett:PressureEffectsMicrobialProc}, which has also been found for biological
membranes from several deep-sea fish tissues~\cite{Cossins:FattyAcidComp}. This adaption has an influence on the physical behavior of a biological membrane
such as the melting transition behavior. 

Melting transitions from a solid ordered
(\textit{so}; in the literature also called \textit{gel}) to a liquid disordered
(\textit{ld}; or named as \textit{fluid}) phase are present in artificial and biological
membranes  \cite{Hinz:dscbilayer, mabrey:dsclipid, Jackson:EcoliDSC, Steim:dscmembr, Reinert:dsc_memlipid,
melchior:transbiomembr}. In the presence of cholesterol also \textit{liquid
ordered} phases develop~\cite{Ipsen:soldlo}. Considering single lipid membranes one finds that a
higher degree of saturation or longer fatty acids result in higher melting
temperatures \cite{Koynova:phasetransPC}. Therefore, one would expect a change of the melting
behavior of biological membranes in dependence of growth temperature or
hydrostatic pressure and therewith in dependence of lipid synthesis. Indeed,
this has been demonstrated for the prominent example of the inner membrane of
Escherichia coli~\cite{Heimburg:genanesth}.

The dependence of lipid synthesis by outer physical parameters such as
temperature or hydrostatic pressure was termed ``homeoviscous
adaption''~\cite{Sinensky:homeoadap, Cossins:MolOrdDeapSea}. The underlying
idea was that for a proper cell function the membrane needs to provide a
certain viscosity. In this context  the widely
used term ``fluidity'' was introduced in the literature. It should, however, be noted that this term lacks a
clear definition. It was also argued  against a role of ``fluidity''~\cite{Lee:lipidsandtheireffects}. An alternative role of the adaption
of lipid synthesis on outer physical parameters is to ensure a
control of the heterogeneity of the lateral membrane structure \cite{Sackmann:triggprocmembrstruct}. This idea has  obtained special
attention within the discussion about ``rafts'' \cite{Brown:rafts}. 

Artificial lipid membranes consisting of one, two or three lipid components
are suitable systems to study the general physical behavior of lipid phase
transitions in more complex lipid systems . In
the phase coexistence regime domains with a size ranging from a few nanometers
to several micrometers form in model systems and can be observed by a large
variety of experimental techniques, such as Fluorescence Resonance Energy
Transfer, Atomic Force Microscopy and Confocal Fluorescence Microscopy
\cite{Feigenson:ternphasediag, Leidy:latorgdomform, korl:cfmfcs,
Bagatolli:phasetransguv, Fidorra:Absenceoffluidord}. 

Already since the early 1970s the idea that domain formation processes can
influence protein activity and function was expressed
\cite{Chapman:liqcrypropphoslip, traeuble:schaltprozmemb,
Sackmann:triggprocmembrstruct}. In several publications the existence of domains
have been shown to provide a mechanism to control biochemical
reaction cascades \cite{Melo:domainconnectionreactions, Vaz:phasetopol,
Thompson:domainstrucreaction, Hinderliter:controlsigntrans,
Salinas:changesenzymeactivity}. Evidence that lipid phase transitions
and domain coexistence trigger enzyme activity has been found for various systems. For example, the enzyme
phospholipase $A_2$ which hydrolyzes the \textit{sn-2} bond of phospholipids,
producing a free fatty acid and a lysolipid, is strongest in activity when
the lipid membrane is in the lipid phase transition regime
\cite{OpdenKamp:phospholipase, Lichtenberg:phospholipase,
Grainger:phospholipase}. The activity of the special class of phospholipase
A$_2$ type II A has been demonstrated to
be related to an enrichment of anionic lipids into \textit{ld} domains
\cite{Leidy:DomainIndAct}. In this case, a different partition behavior of the anionic lipids into \textit{ld} and \textit{lo}
domains led to a local concentration higher than the threshold
value for enzyme activity. Changes in the opening times and probabilities of
calcium channels reconstituted in POPE:POPC membrane of various ratios have
been interpreted to be due to the coexistence of solid ordered and liquid
disordered phases~\cite{Cannon:RegCalciumChannel}.  A further example is the protein kinase C (PKC) which
modifies proteins by chemically adding phosphate groups. It has been discussed
that lateral heterogeneities of the lipid membrane control the activation of
this enzyme \cite{Bolen:unstapkcactivation, Dibble:laterheteroactpkc}. In the
latter study it has also been pointed out that domains enriched in
dioleoylglycerol lipids play a crucial role, but also a linear relationship
between PKC activity and phosphatidylserine content has been found
\cite{Orr:PKC_phosphserin}. Further communications can be retrieved where a
connection between enzyme activity and the presence of specific lipids and
structure are made. This is e.g. also the case for the calcium
ATPase~\cite{Tang:lipstructCaATP}.

\pagestyle{myheadings}
\markboth{Fidorra et al.}{Melting of Single Lipid Components}

The publications cited above are only a small fraction of many in that
respect, but they already show that it is important not only to study domain 
formation, but also to investigate the fraction of
the single lipid species in lipid membranes showing lateral inhomogeneity due
to phase separation. The single lipid species will occur in different amounts
in the different domains that form in a phase transition regime
\cite{Lee:phasedia}. The work presented here concentrates on this for the
case of a binary lipid mixture composed of \textit{DMPC} and
\textit{DSPC}. These two phospholipids display melting transition events with
transition midpoint
temperatures of $23.6 \degree $C or $54.4 \degree $C, respectively
\cite{Koynova:phasetransPC}. In our study we used a deuterated \textit{DMPC}
derivate whose phase transition midpoint temperature lies at around $19-20 \degree C$ \cite{GuardFriar:DeutPhospho,Wang:thermdeutbiolmolec}.

For the description of phase separation in lipid membranes various
theoretical approaches have been applied. This includes ideal and regular
solution theory, mean field, but also numerical approaches such as Monte Carlo
(MC) simulations \cite{Lee:phasedia, marcelja:bimemb, pink:ramansctpc, sugar:MCSIsing, sugar:TwoCompLipid}. Ideal and regular solution theories allow the understanding of
phase diagrams. In the ideal solution theory complete
miscibility in the solid ordered and liquid disordered phase is assumed. The
regular solution theory takes a step further and regards a non-ideal mixing in
the solid ordered phase. These considerations allow the construction of phase
diagrams. Phase diagrams are used to map the phase behavior of lipid membranes
in dependence of e.g. temperature and lipid composition. The theoretically derived lever rule provides the possibility to
determine the ratio of disordered and ordered lipids in dependence of
temperature and lipid composition of binary lipid mixtures and can also be
applied to experimentally measured phase diagrams \cite{Lee:phasedia}. Ideal
or regular solution theory are, however, based on simplifying
assumptions. This does not only include an ideal mixing in the liquid phase,
but also the assumption of phase separation under all conditions in the
melting regime. This is, however, not the case
\cite{michonova:statesep}. Further, fluctuations which are enhanced in the
melting transition regime are present \cite{heim:mech98,seeger:fluct05}. This
is where  MC
simulations analyzing simple models of lipid chain melting enter. They allow the consideration of interactions between different lipid
species and varying chain state. The formation of lipid domains and
phases and the consideration of fluctuations is a direct consequence of the
thermodynamics of these systems which is included in these models. Therefore,
these
simulations can describe lipid membrane properties in the vicinity of lipid phase transitions, such as changes in
membrane permeability \cite{Cruzeiro:permeability}, fluctuations in enthalpy
\cite{seeger:fluct05}  and diffusion processes  \cite{hac:dmdsdiff,
Sugar:diffusion} well. 

In this
study we determined the amount of the single lipid species in the
different domains of binary \textit{DMPC} and \textit{DSPC} mixtures by means of MC
simulations and compared the results obtained to experimental data measured
by Fourier Transform InfraRed spectroscopy (FTIR).

Among other techniques FTIR is well suited to investigate the melting
behavior of lipid membranes. In these studies the temperature dependence of
various vibrational modes due to structural changes provide a versatile tool in
following phase transitions \cite{Tamm:RevIR}. Articles have been published on experiments in which
FTIR was used to measure nanoscale domain size \cite{Mendelsohn:IRspect} and
chain order parameters \cite{Reinl:ChangPhysProp}. 

In our study the hydrogen atoms of \textit{DMPC} were replaced by deuterium atoms. Due
to this the melting of the \textit{DMPC} lipid chains could be separated from the \textit{DSPC}
lipid chain melting, as the $CH_2$ and $CD_2$ symmetric and asymmetric stretch
vibrations differ in frequency due to the heavier deuterium atoms. This allowed
us to compare the melting of single components with MC simulation results
which contain information of the disordered chain ratio of single lipid
components. MC simulations are able to describe experimental heat capacity
profiles well. Therefore, they provide a means to couple the two
different experimental approaches. 

The experimental absorption spectra were evaluated using two
methods. One of the methods probed the temperature dependence of the position
of the absorption maxima of the antisymmetric $CD_2$ and symmetric $CH_2$
stretch vibrations. The other one evaluated peak area changes of
difference spectra (\textit{Difference Spectra method}). Simulation
results on the fraction of disordered chains of the single lipid species
agreed well with the data found using the latter method. FTIR allows to
distinguish the melting of the single lipid species and it probes the
microscopic behavior. DSC probes the bulk
behavior and macroscopic observables. The Monte Carlo simulations were used as
an approach to combine the informations of the two experimental techniques. DSC, MC and FTIR are a potent combination to investigate the melting behavior of lipid membranes.


\section*{Materials and Methods}

{\noindent\textbf{Sample Preparation:}} 

 The lipids 1,2-dimyristoyl-d54-sn-glycero-3-phosphocholine (\textit{DMPC-d54}) and 1,2-distearoyl-sn-glycero-3-phosphocholine (\textit{DSPC}) were purchased from Avanti Polar Lipids (Alabaster, USA).
Samples were prepared from lipid stock solution in chloroform. These
were prepared in chloroform from lipid powder as delivered from the vendor
without further purification. To prepare lipid samples in buffer, chloroform
from the desired amount of lipid stock solution mixture was evaporated under a
stream of nitrogen and the sample was kept under vacuum for several hours
afterwards. After this procedure the dried lipid film was hydrated with
preheated ultrapure water at a temperature of 60\degree C, which is higher
than the melting temperature of the higher melting lipid component \textit{DSPC}
($54.4 \degree$ C, \cite{Koynova:phasetransPC}) and kept stirring at this temperature for
about 30 minutes. In calorimetric experiments a lipid concentration of $10\:
mM$ was used. The Fourier Transform Infrared spectroscopy measurements required
concentrations of $65\: mM$. In order to facilitate lipid membrane hydration
 under these high concentrations the samples were subjected to several freeze-thaw cycles.     

\

{\noindent \textbf{DSC experiments:} }

 Differential Scanning Calorimetry (DSC) experiments were performed on a VP-DSC
(MicroCal, Northampton, MA, USA) with a scan rate of 15\degree C/h
(DMPC-d54:DSPC $50:50$) or of 5\degree C/h (DMPC-d54) per hour in upscan
direction. The samples were equilibrated at the starting temperature for 30
minutes.

\ 

{\noindent\textbf{Fourier Transform InfraRed spectroscopy experiments:} }

 Fourier Transform InfraRed (FTIR) spectroscopy experiments were performed on
a Vertex70 (Bruker Optics, Bremen, Germany) equipped with a
mercury-cadmium-telluride (MCT) detector.  The spectrometer was flushed with nitrogen
starting about 30 minutes before the experiments. Temperature control was done
by a self built sample holder connected to a water bath (DC30-K20,
ThermoHaake, Karlsruhe, Germany). The scan rate for the temperature
ramp was set to 20\degree C/h. The temperature in
the sample corresponding to each temperature point set with the water bath was
appointed in a reference scan. Background spectra were acquired at
  each temperature point after lifting the sample compartment out of the beam
  path. We were interested in the IR absorption due
to the symmetric and asymmetric stretch vibrational modes of the $CH_2$ and
$CD_2$ groups of the lipid fatty acid chains. Example absorption maxima of the $CH_2$
symmetric and asymmetric stretch vibrations at different temperatures are displayed in the left panel of 
fig.~\ref{fig:fig2}. 

Two methods to analyze the FTIR spectroscopy data were used. In a first approach FTIR spectroscopy data was processed with a self-written plugin for the IGOR
software package (WaveMetrics, Lake Oswego, OR, USA). The absorption
  peaks were fitted with a polynomial function. Then the peak position was
  determined by calculating the root of the fit.

For the second protocol the peak area of the $CH_2$ or respectively of
the $CD_2$ stretch vibrations was normalized to a value of $1\times cm^{-1}$. A reference
spectrum was chosen which was taken at a temperature below the melting
transition regime. This spectrum was subtracted from all other spectra which
were recorded at different temperatures (see the right panel of
fig.~\ref{fig:fig2}). Then the peak area located at the positive axis of
ordinates was calculated and plotted as a function of temperature. There was
no difference in surveying both, symmetric and antisymmetric modes or the
single ones. Further, we shall call this method the \textit{Difference Spectra
method}. 

\ 

{\noindent\textbf{Model:}}

We have already applied an Ising-like model to the study of
diffusion \cite{hac:dmdsdiff}, fluctuation \cite{seeger:fluct05} and
relaxation properties \cite{seeger:binary_relaxproc} of \textit{DMPC:DSPC} lipid
mixtures. In this paper we used this model to describe FTIR measurements on
 mixtures of \textit{DMPC-d54} and \textit{DSPC} and linked these with DSC measurements. 

The model has been presented in detail elsewhere \cite{hac:dmdsdiff,
  sugar:TwoCompLipid}. In short, the model is based on the assumption that the
lipid chain states can be described by assigning  either an ordered or
disordered state to the fatty acid lipid chains. Lipid chains are arranged on
a triangular lattice and only nearest neighbor interactions  are
included. Lipid chains might change the degree of order or their positions
during the simulation. The model is evaluated with Monte Carlo (MC) 
simulations in which the free energy difference of an old and a trial
configuration is sufficient for decision making of the acceptance of a new
configuration. The free energy of the system can be divided into a
configuration dependent and an independent part. The configuration dependent
contribution equals~\cite{sugar:TwoCompLipid}:

\begin{eqnarray}
\nonumber \Delta G(\mathbf{S})&=& N_1^d(\Delta H_1 - T \Delta
S_1)+N_2^d(\Delta H_2- \Delta S_2)\\ & + &
N_{11}^{od}\omega_{11}^{od}+N_{12}^{oo}\omega_{12}^{oo}+N_{12}^{od}\omega_{12}^{od}\\
\nonumber &+& N_{12}^{dd}\omega_{12}^{dd}+N_{12}^{do}\omega_{12}^{do}+N_{22}^{od}\omega_{22}^{od},
\label{eq:Gibbsfreeenergy_eq1}
\end{eqnarray}  

where $\Delta H_i$ and $\Delta S_i$ are the enthalpy and respective entropy differences of
the disordered to the ordered state of \textit{DMPC} ($i=1$) and \textit{DSPC} ($i=2$) lipid
chains. $\omega_{ij}^{mn}$ are interaction parameters of unlike nearest
neighbors and $N_{ij}^{mn}$ the corresponding number of the unlike nearest
neighbor contacts with $(i\neq j)$ and $(m\neq n)$. $m$ and $n$ denote the
state of a chain which is either ordered ($o$) or disordered ($d$). Enthalpy,
entropy and nearest neighbor interaction parameters can be obtained from
experimentally determined heat capacity profiles. Previously, we have determined
the parameters needed for the \textit{DMPC:DSPC} system \cite{hac:dmdsdiff}. In this
paper, however, FTIR experiments were performed with \textit{DMPC-d54} and
\textit{DSPC} mixtures. In the following it was assumed that the unlike species
interaction parameters did not change from the system \textit{DMPC:DSPC} to
\textit{DMPC-d54:DSPC}. The heat capacity profile of the pure
\textit{DMPC-d54} was measured to obtain its enthalpy and entropy changes
during the melting transition and the cooperativity parameter
$\omega_{11}^{od}$ (\textit{data not shown}). The remaining parameters were
taken from a recent publication \cite{hac:dmdsdiff}. All  values are listed
in table ~\ref{tab:parameter}. 

\begin{center}
\begin{table}[h!]
\linespread{1.3}
\footnotesize
\begin{center}
	\begin{tabular}[t]{l l r r l l}
\hline 
$T_{m,1}$ = 19.7&\degree C & & &   $\omega_{11}^{od}$ = 1267 &J/mol \\
$T_{m,2}$ = 54.8&\degree C & & &  $\omega_{22}^{od}$ = 1474 &J/mol \\
$\Delta H_1$ = 14500& J/mol & & &        $\omega_{12}^{oo}$ = 607 &J/mol\\ 
$\Delta S_1$ = 49.5  &J/(mol \degree C) & & &       $\omega_{12}^{dd}$ = 251
&J/mol \\
$\Delta H_2$ = 25370 &J/mol & & &     $\omega_{12}^{od}$ = 1548 &J/mol \\
$\Delta S_2$ = 77.36 &J/(mol \degree C) & & & $\omega_{12}^{do}$ = 1716 &J/mol
\\ \hline
\end{tabular}
\end{center}
\linespread{1}
\selectfont
\caption[Parameter Values in the Simulation]{\footnotesize{Parameter values
    used in the Monte Carlo simulations. They were determined from
    experimentally determined heat capacity profiles \cite{hac:dmdsdiff}. All
    numbers are given per lipid chain. The indices \textit{o} and \textit{d}
    stand for ordered and disordered, respectively. \textit{1} and \textit{2}
    index \textit{DMPC-d54} and \textit{DSPC}.}}
\label{tab:parameter}
\end{table}
\end{center}

The simulations were done on matrix sizes $60\times60$ (\textit{DMPC-d54:DSPC} 100:0, 70:30, 60:40, 50:50,
40:60, 30:70), $100\times 100$ (90:10, 80:20, 20:80, 10:90) and
$350\times 350$ (\textit{DSPC}) \cite{sugar:TwoCompLipid}.  The system was equilibrated for
$5000$ or around the heat capacity maxima for $10000$ MC
cycles. Simulations were conducted over $50000$ or $100000$ MC cycles,
respectively.  


\section*{Results}

In this paper we discuss melting processes in binary lipid
mixtures. Thereby, we focus on the melting of the two single
components independently. Experimental FTIR and DSC measurements were accompanied by numerical
simulations and an analysis of the corresponding phase diagram.   

\begin{figure}[b!]
    \includegraphics[width=8cm]{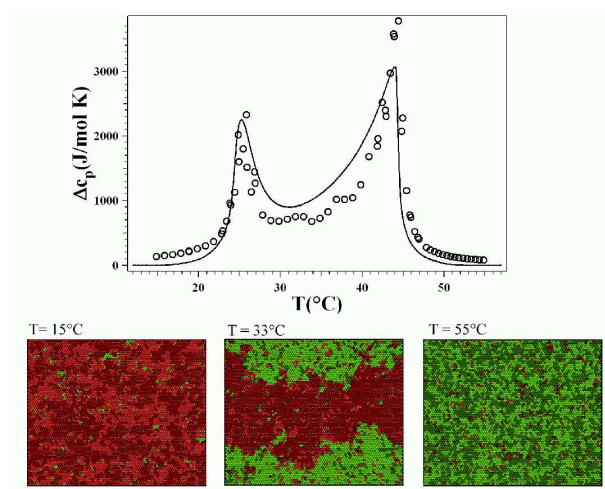}
    \parbox[c]{8cm}{\caption{\textit{Enthalpy and entropy changes and cooperativity parameters of the
single DMPC-d54 and DSPC lipid membranes were determined from heat capacity
profiles.  The remaining four unlike species nearest neighbor
interaction parameters of DMPC-d54:DSPC mixtures were assumed to equal the
ones of a DMPC:DSPC system \cite{hac:dmdsdiff}. Simulating an equimolar
mixture of DMPC-d54 and DSPC gives a melting profile
(markers) which describes well the measured heat capacity profile
(solid curve). Below three representative snapshots from simulations
of the same mixture with a matrix size of $80\times 80$ lipid chains are shown. The red and green colors represent ordered or
disordered lipid chains, respectively. The lighter colors correspond to the
DMPC-d54 lipid chains and the darker ones to the DSPC
lipid chains. The snapshots were taken at three different temperatures. At a
temperature of $15\ \degree$ C the membrane is in the solid ordered
phase while at a temperature of $55\ \degree$ C it is in the liquid
disordered phase. In both cases lipids do not mix ideally, but lateral
hetereogeneities are present. Two macroscopic phases are present at a
temperature of $33\ \degree$ C.}
    \label{fig:fig1}}}
\end{figure}

MC simulations were used to evaluate a simple model based on the
assumption of two lipid chain states, the consideration of nearest
neighbor interactions only and the application of a hexagonal lattice. This
model has already been shown to  describe melting processes in
\textit{DMPC:DSPC} mixtures successfully \cite{sugar:TwoCompLipid, hac:dmdsdiff, seeger:fluct05}. The
evaluation needs the determination of ten parameters. These are twice enthalpy
and entropy differences between the two disordered and ordered chain states
and six unlike nearest neighbor interaction parameters. The enthalpy and
entropy changes can be  deduced easily from heat capacity measurements of the
single one component systems. Two interaction parameters describe the cooperativity,
meaning the width of these transitions~\cite{Heimburg:MCDSC,
  Heimburg:therminteractprotlip}. The four parameters  remaining need to
be deduced from comparisons with $c_p$-profiles obtained for different lipid
compositions. Previously, we have determined the parameters for a
\textit{DMPC:DSPC} system \cite{hac:dmdsdiff}. In this study, however, we
exchanged the \textit{DMPC} lipids with deuterated \textit{DMPC (DMPC-d54)}
which gives heat capacity profiles which are broadened and shifted by
$-3.9\degree C$ to lower temperatures in comparison to the nondeuterated
\textit{DMPC} (\textit{data not shown} and see \cite{GuardFriar:DeutPhospho, Wang:thermdeutbiolmolec}). This means that we had to determine the
enthalpy and entropy changes and the cooperativity parameter of pure
\textit{DMPC-d54}. In principle using this model one needs to reestablish all
parameters for any  binary lipid system. In this work, however, we assumed that the unlike nearest
neighbor interaction parameters do not change by the replacing of
\textit{DMPC} with its deuterated analog and we adopted them from  previous
simulations of the nondeuterated \textit{DMPC:DSPC} system \cite{hac:dmdsdiff}. To
validate this,  DSC measurements on an equimolar \textit{DMPC:DSPC} mixture
were performed and they were compared  to results simulated. We found that the
values calculated describe  the melting profile measured well (see
fig.~\ref{fig:fig1}). Therefore, we were convinced that it was feasible to
further use the model with the interaction parameters given in table~\ref{tab:parameter}. 

The further scope of this paper was to compare
the temperature dependence of the simulated disordered chain ratios of
\textit{DMPC-d54} and \textit{DSPC} lipids with the experimental analysis of
the melting processes using FTIR spectroscopy. This indirectly allowed us to
compare the experimental results obtained with DSC and FTIR.

\begin{figure*}[htb]
\begin{center}
    \includegraphics[width=15cm]{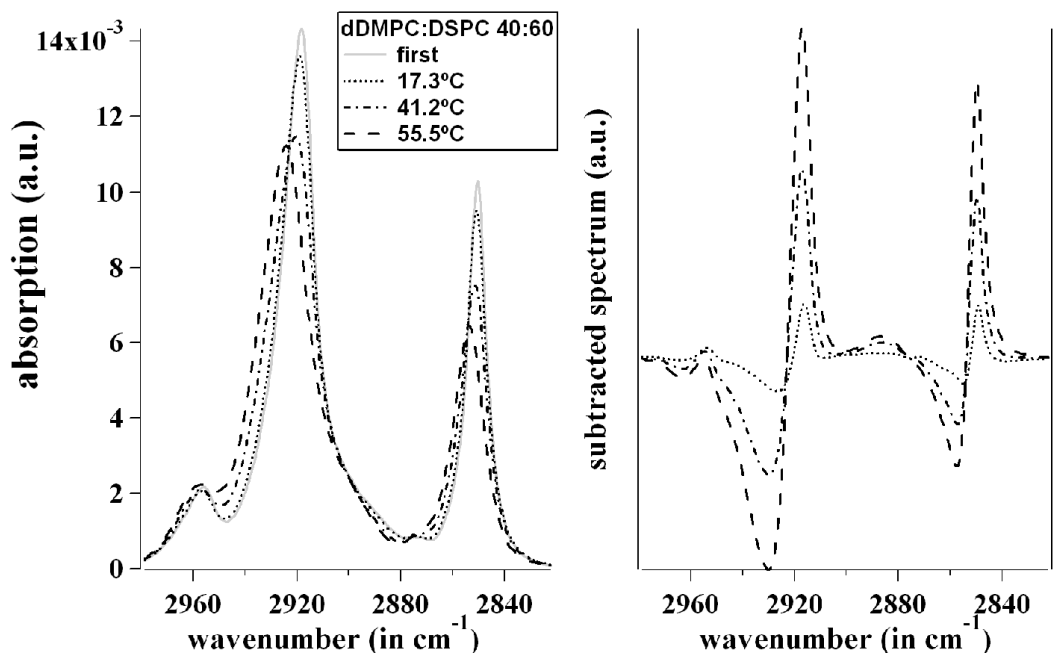}
    \parbox[c]{16cm}{\caption{\textit{The IR absorption spectrum contains different absorption maxima originating
from varying vibrations. In the \textit{left} panel the asymmetric and the symmetric
$CH_2$ stretch vibrations  as measured in a sample of \textit{DMPC-d54:DSPC 40:60} at
different temperatures are displayed. The absorption maxima of these are
temperature dependent and lie around $2852\: cm^{-1}$ (symmetric) and at
$2925\: cm^{-1}$ (asymmetric). The asymmetric and symmetric $CD_2$
stretch vibrations are shifted to lower wavenumbers due to the heavier
deuterium atoms (\textit{data not shown}). In previous studies from other labs
the melting processes were investigated following the temperature dependence
of the absorption maxima. In this study, however, the
melting process was also studied by defining a  reference spectra and subtracting
this one from all other spectra where in each cases the area was normalized to
$1\times
cm^{-1}$ (\textit{Difference Spectra method}). Example results of the difference spectra of the left panel are
displayed in the \textit{right} panel. As a measure of the progress of the
melting process the area of one of the peaks was taken. }
\label{fig:fig2}}}
    \end{center}
\end{figure*}

In FTIR spectroscopy the absorption of infrared light due to
vibrational degrees of freedom in the sample is studied. In lipid suspensions
different vibrational modes of the lipids contribute to the absorption
spectrum. Prominent examples used in our study are  $CH_2$ or $CD_2$ (in the
case of the deuterated \textit{DMPC} lipids)  stretch vibrations of the fatty
acid lipid chains. The shape of the absorption band is temperature dependent
as can be seen in the left panel of fig.~\ref{fig:fig2}. Example spectra of the
symmetric and asymmetric $CH_2$ vibrational spectra are given for different
temperatures of a mixture of \textit{DMPC-d54:DSPC 40:60}. Deuterium atoms are
heavier than hydrogen atoms. This leads to a shift of the $CD_2$ stretch
vibrations to lower wavenumbers and allows to distinguish the two
different lipid species in a binary lipid mixture.\\

\

\textbf{Comparison of FTIR Results and Monte Carlo Simulations}

\

In an investigation of structural changes in cytochrome c Heimburg and Marsh
~\cite{Heimburg:CyctC} focused on temperature dependent changes of
difference spectra. This was motivated by the idea that the corresponding amide
I band is a convolution of a variety of different bands. The temperature
dependence of each of the single bands differs and a complex temperature
dependent behavior is present. Therefore, we decided not only to
investigate the temperature dependence of the absorption maxima as done in the
literature previously \cite{Dluhy:FTIRDPPCDMPC, Mantsch:FTIR,
Leidy:latorgdomform}, but also to use a \textit{Difference Spectra method}
too. Further
we studied the peak area of these difference spectra as a function of temperature. 

In the left panel of fig.~\ref{fig:fig2} example spectra of the $CH_2$ stretch
vibrations are given as a function of temperature of a \textit{DMPC-d54:DSPC 40:60} mixture.  Normalizing the area of the
spectrum to $1\times cm^{-1}$ and defining a reference spectrum which lies at
a temperature below the melting transition regime, difference
spectra were calculated. Representative examples are displayed in the right
panel of fig.~\ref{fig:fig2}. To follow the melting transition the change in
difference spectra peak area of the $CH_2$ or the $CD_2$ symmetric, asymmetric
or both vibrational modes can be
monitored. Our experiments showed that there is no difference between these
three approaches (\textit{data not shown}). 

\begin{figure*}[htb]
\begin{center}
    \includegraphics[width=15cm]{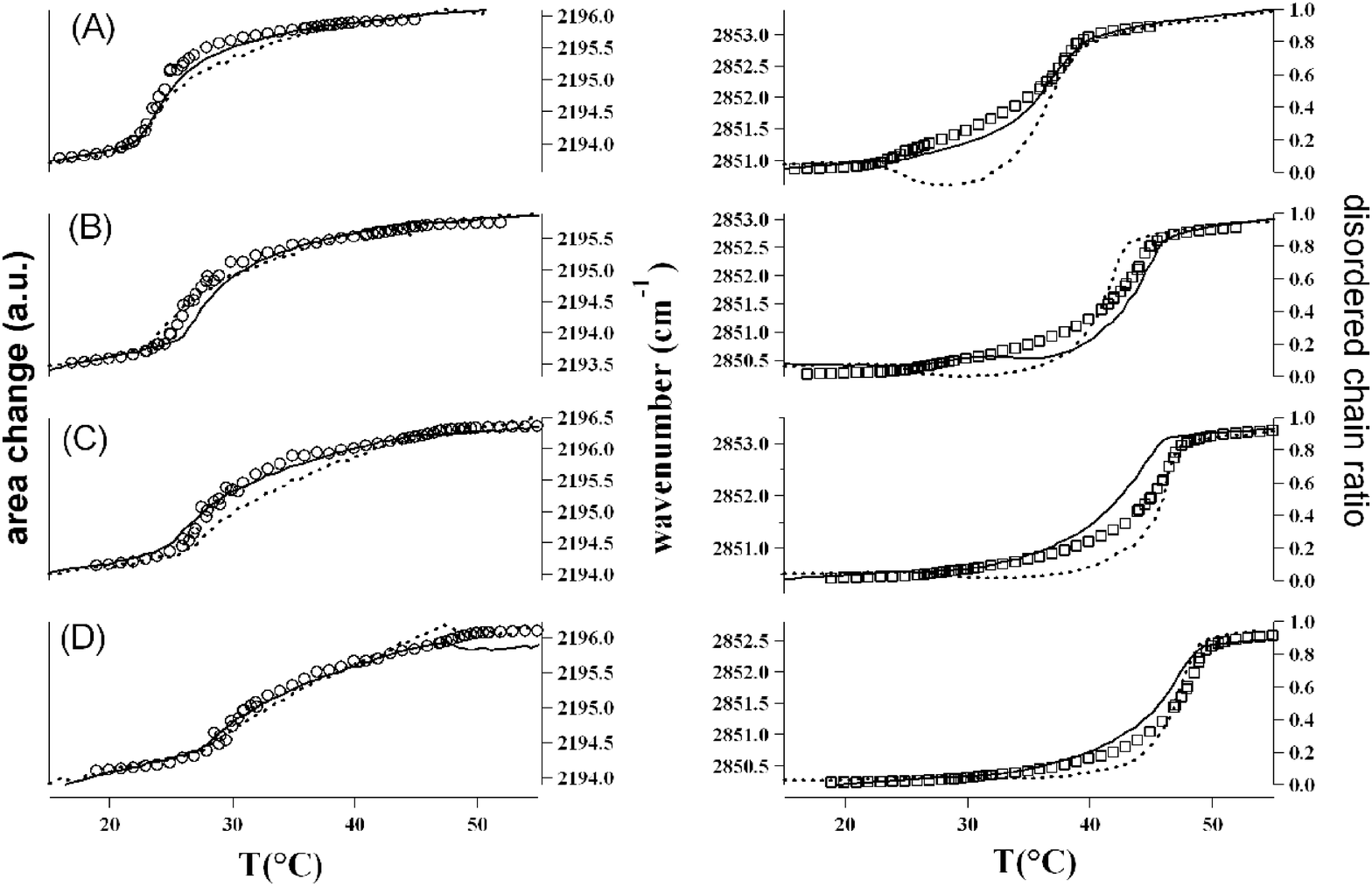}
    \parbox[c]{16cm}{\caption{\textit{The FTIR results are compared with the Monte Carlo simulations. The data on four different \textit{DMPC-d54:DSPC} ratios \textit{70:30} (A),
\textit{50:50} (B), \textit{40:60} (C) and \textit{30:70} (D) are
displayed. In each panel the disordered chain ratios obtained from the
simulations (open circles (\textit{DMPC-d54}) and open
squares (\textit{DSPC})) are compared with the changes of the peak area (solid
curves) and the evolution of the position of the absorption maximum (dashed
curves). In the latter case the antisymmetric $CD_2$
(\textit{DMPC-d54}) and the symmetric $CH_2$ stretch vibrations were followed. The panels on the left side depict the melting process of the
deuterated \textit{DMPC-d54} lipids and the right ones the \textit{DSPC} lipid
melting. The Monte Carlo simulations allow to combine both the FTIR and the
calorimetric results. They provide a mean to relate the macroscopic with the
microscopic behavior of the binary lipid membrane melting.}
\label{fig:fig3}}}
    \end{center}
\end{figure*}

The temperature dependent evolution of the difference spectra peak areas
monitored are given in fig.~\ref{fig:fig3} as solid curves for four different
\textit{DMPC-d54:DSPC} mixtures: (A) \textit{70:30},  (B) \textit{50:50},  (C)
\textit{40:60} and  (D) \textit{30:70}. The panels on the left side depict the
results related to the deuterated \textit{DMPC-d54} lipids and the right ones
give the data belonging to the \textit{DSPC} lipids. The data obtained is also compared to simulation results of the
ratio of disordered lipid chains for each of the single species. In
fig.~\ref{fig:fig3} these are given as open circles (\textit{DMPC-d54}) or
squares (\textit{DSPC}). In all instances we found an increase of the
area with higher temperatures. In all cases a good agreement between the
simulated and measured values is present. The deviations of the data
simulated from the measured ones are smaller in the cases of the
\textit{DMPC-d54} lipids. This might be due to the fact that the determination
of the model parameters needed is imperfect. More extended and time-consuming
simulations might improve the data, too.

As already mentioned above, previously melting processes of lipid membranes
 investigated by FTIR were analyzed monitoring the temperature dependence of
 the position of the absorption maximum of either the symmetric or asymmetric
 stretch vibrations of $CH_2$ and $CD_2$. We analyzed the temperature
 dependence of the absorption maxima for several \textit{DMPC-d54:DSPC}
 mixtures as earlier done by Leidy et al. \cite{Leidy:latorgdomform}, too. The results
 obtained are given by the dashed curves in fig.~\ref{fig:fig3}. 

 We confirmed what has been reported by Leidy et al. 
\cite{Leidy:latorgdomform} previously. There are instances when the
positions of the maxima of the symmetric stretch vibrations of the \textit{DSPC} lipids do not increase with higher
temperatures, but that for certain lipid mixture fractions they decrease, approach a minimum and then they start to become larger
again. The minimum wavenumber is in all cases lower than the wavenumbers determined at lower temperatures. This behavior is in disagreement with the evolution of the difference spectra
peak area and the MC simulation results.

\begin{figure*}[htb]
\begin{center}
    \includegraphics[width=15cm]{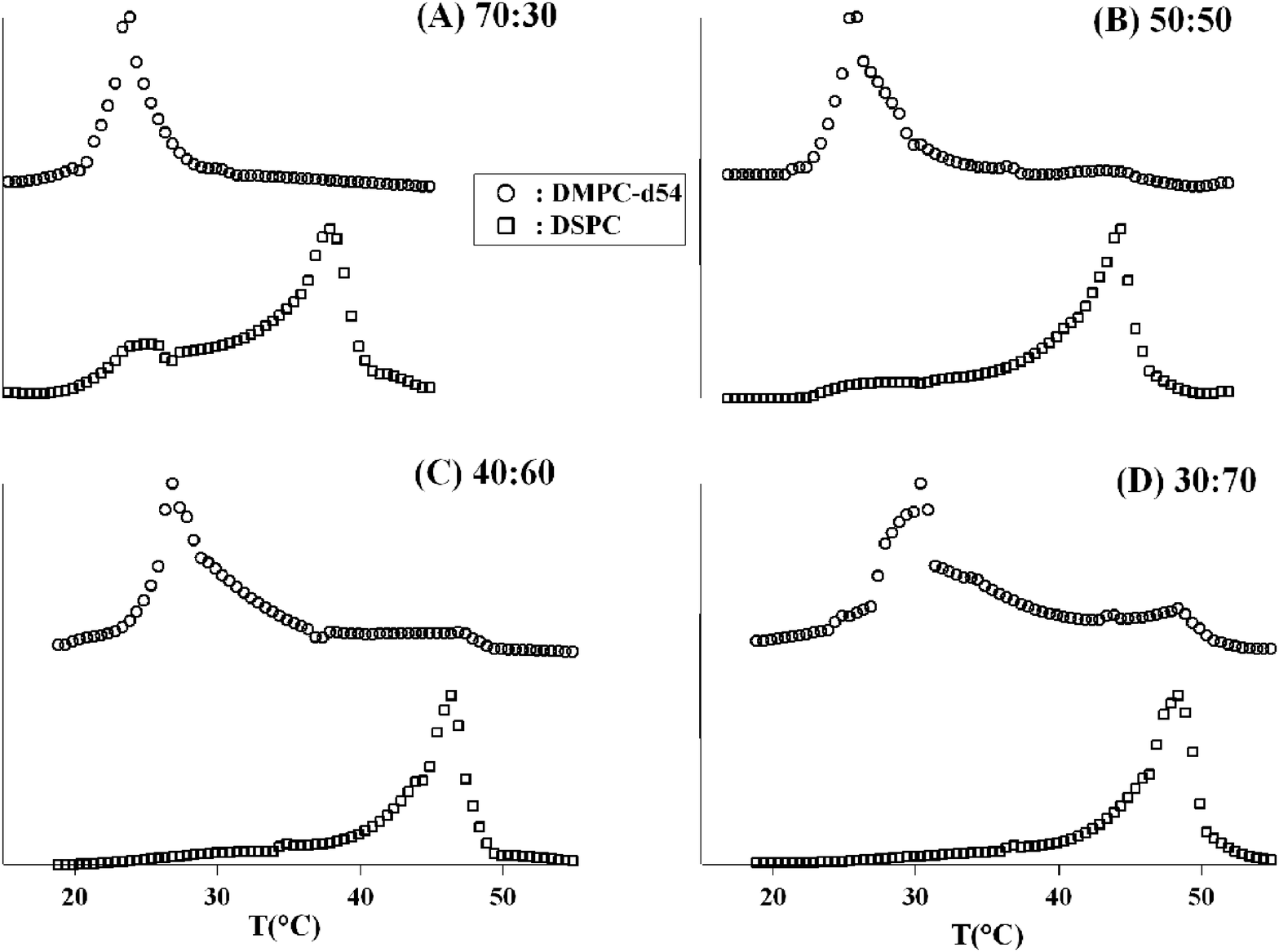}
    \parbox[c]{16cm}{\caption{\textit{The derivatives of the curves of the fraction of disordered \textit{DMPC-d54}
and \textit{DSPC} lipid chains as shown in fig.~\ref{fig:fig3} are displayed
in this figure. In detail these are the mixtures \textit{70:30} (A), \textit{50:50} (B), \textit{40:60}
(C) and \textit{30:70} (D). The derivatives belonging to \textit{DMPC-d54} are
indicated by open circles, while the ones of \textit{DSPC} by open squares. A
maximum in the derivative means that at this temperature fluctuations
are enhanced. }
\label{fig:fig4}}}
    \end{center}
\end{figure*}

It is well known that macroscopic, but also
local fluctuations of various membrane properties are enhanced in the vicinity
of lipid membrane melting transitions
\cite{heim:mech98,seeger:fluct05}. The determination of the derivatives of the data of
the curves from fig.~\ref{fig:fig3} as done by Leidy and collaborators\cite{Leidy:latorgdomform}
provides a measure of the strength of fluctuations with respect to the single
lipid components. In fig.~\ref{fig:fig4} these derivatives are shown for the
data simulated. The higher the value of the derivative the stronger the
fluctuations. \textit{DMPC-d54} melts at lower temperatures than
\textit{DSPC}. This is why the corresponding derivatives in the left panels of
fig.~\ref{fig:fig4} have maxima at lower temperatures. This temperature,
however, depends on the molar fraction of \textit{DSPC} lipids and is
higher with an increased  \textit{DSPC} lipid fraction. This is similar for
the \textit{DSPC} lipids only with the difference that the presence of
\textit{DMPC-d54} lowers the temperatures at which fluctuations of the
\textit{DSPC} lipid chains are the strongest. In this context it should be
noted that the component which is in the minority always displays two
maxima. One of the maxima is a local maximum at which fluctuations are
enhanced, but they are lower than at the global maximum. The local maximum is
found at a temperature at which the other component displays its
strongest fluctuations. At an equimolar ratio of \textit{DMPC-d54:DSPC} this is true
for both components. The melting transition of the single lipid components
happens over a broad temperature regime.  

\

\textbf{The Phase Diagram}

\

\begin{figure}[b!]
\begin{center}
    \includegraphics[width=8cm]{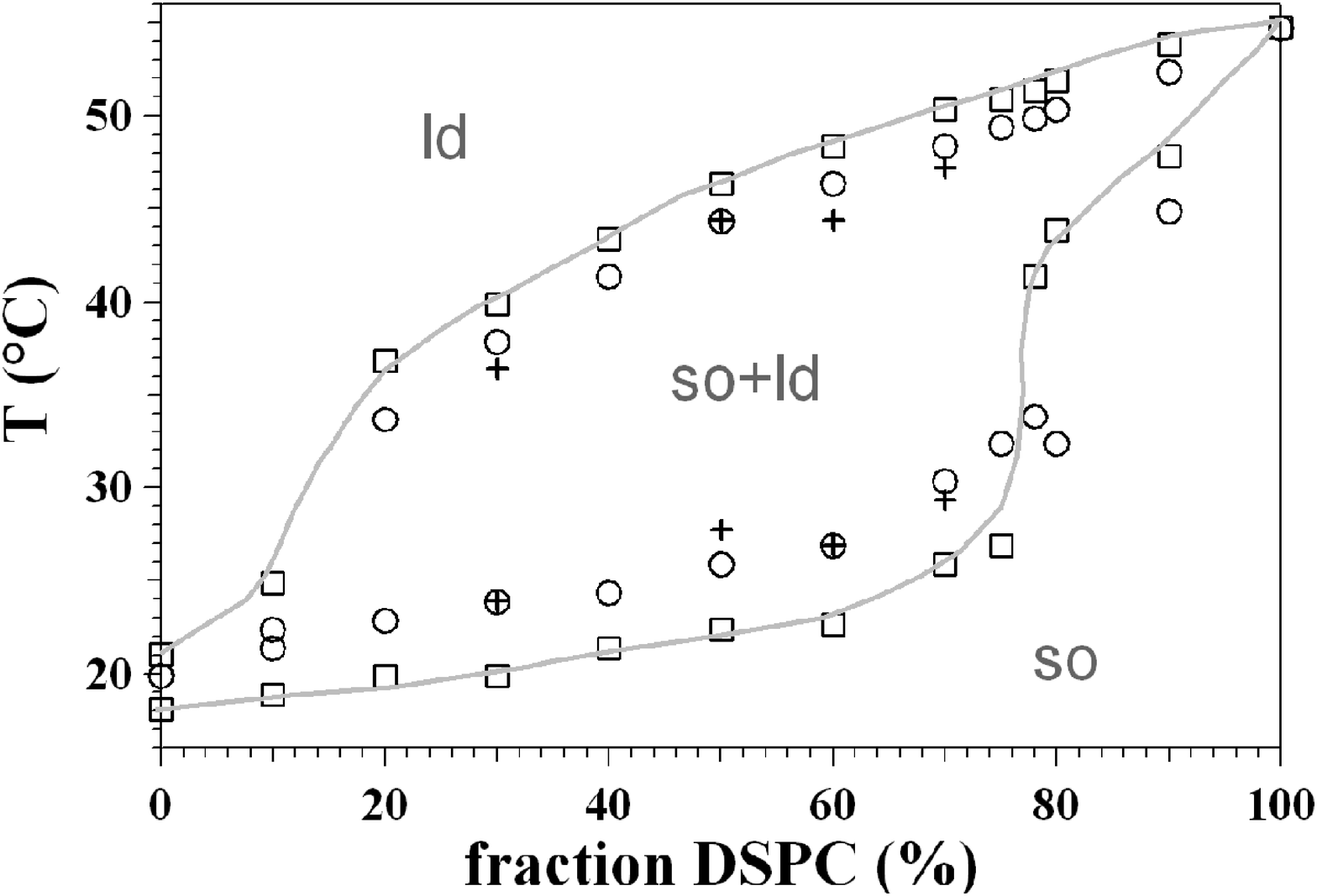}
    \parbox[c]{8cm}{\caption{\textit{ Phase diagrams allow to obtain an understanding of the phase behavior of lipid
mixtures in dependence of lipid ratios, temperature, pH or other outer
parameters. In this study binary lipid mixtures were used to study melting
transitions in dependence of temperature. Therefore, a phase
diagram with the fraction of \textit{DSPC} lipids as the abscissa and temperature as
the ordinate was constructed. The open squares and the grey lines represent the lower and upper temperature
limits of the melting process as determined from heat capacity
profiles simulated using the tangent method. Further the peak positions of
the derivatives of the  disordered chain fractions simulated (open circles) and
of the experimentally determined area changes (crosses) were analyzed. The simulated peaks
determined from the derivatives agree with the maxima of the heat capacity
profiles (\textit{data not shown})}
\label{fig:fig5}}}
    \end{center}
\end{figure}

Phase diagrams are generally constructed to easily access the phase separation
behavior of complex lipid mixtures \cite{Lee:phasedia}. In our case a phase
diagram depending on lipid composition and
temperature was constructed. A phenomenological method to construct these is in aligning
tangents on the lower and upper temperature limits of heat capacity
profiles. This results in the construction of a solidus and a liquidus line as
indicated by open squares and the grey curves in fig.~\ref{fig:fig5}. If the lipid composition is
known,  the physical state of the membrane can be determined. The
data obtained from the lower temperature limit defines the solidus line,
whereas the one from the upper temperature limit gives the liquidus line. Below
the solidus line the lipid membrane is in the \textit{solid ordered} phase,
while above the liquidus line it is in the \textit{liquid disordered}
phase. Therewith, the liquidus and the solidus lines define the phase
coexistence regime too. The three snapshots in fig.~\ref{fig:fig1} provide a visual
impression of this. At temperatures of $15\degree$ C and $55\degree$ C the
lipid membrane is in either the \textit{solid ordered} or \textit{liquid
disordered} phase, respectively. It should, however, be mentioned that the
mixing of the two lipids is in neither of the two cases ideal. The single
lipid species cluster. At a temperature of $33\degree$ C the two phases
coexist and a macroscopic phase separation is present. In an earlier
publication we could show that discussing domain formation processes in lipid
membranes one needs to distinguish microscopic and macroscopic phase
separation and local fluctuations in chain state enter the physical
picture. These fluctuations are enhanced at domain or phase boundaries. Small
domains are also subject to strong fluctuations
\cite{seeger:fluct05}. Analyzing phase diagrams one should note that these
physical effects are in general not considered in its interpretation. Further,
it has been pointed out that the construction of a phase diagram is strictly
speaking only valid if the transition is of first-order
\cite{sugar:twocom_enzym}. The transition of binary lipid mixtures of
\textit{DMPC} and \textit{DSPC} has been claimed to be of second order and
therefore the interpretation of the phase diagram has to be taken with care \cite{michonova:statesep}.

With the determination of the derivatives  of the curves in
fig.~\ref{fig:fig3} we were able to deduce the temperatures at which the
fluctuations of the single components were enhanced. The lower temperature melting
component, in this case \textit{DMPC-d54}, has its maximum at lower
temperatures than the one which melts at higher temperatures
(\textit{DSPC}). These temperatures, however, are higher or respectively lower
than the melting temperatures of the pure single components. The values are
given in fig.~\ref{fig:fig5}. Open circles refer to the values originating from the simulations while the crosses come from the
experiments.  The temperature points determined using the two different
methods to analyze the FTIR absorption spectra are the same (\textit{data not
shown}). The experimental and the simulated data agree well. 

The simulations further allowed us to compare the temperatures at which the
heat capacity curves are in a maximum and the temperatures  at which the
derivatives of the disordered chain ratios display their maxima. We found that these temperatures equal each other (\textit{data not shown}). In some
cases only one heat capacity maximum is resolvable. The analysis of the single components, however, allows to
distinguish the maxima of the melting of single components in all
instances. The maxima of the component which is in a majority then agrees with
the maximum of the heat capacity curve.  

In total, we showed a combined FTIR and MC simulation study. FTIR absorption
spectra were analyzed with two methods. One followed the temperature
dependence of the absorption maxima of the symmetric or asymmetric $CH_2$ or
$CD_2$ stretch vibrations. The other studied temperature dependent changes
of the peak area of  the difference spectra. Monitoring the positions
of band maxima leads to incorrect results, but the simulated data describe
well the curve shapes obtained with the \textit{Difference Spectra method}. Our simulations describe well the evolution of the
area changes. Melting of single components cannot be investigated in DSC. DSC
alone only shows the bulk behavior. In FTIR spectroscopy it is possible to
separate the lipid species individually. Therefore, using MC simulations we
were able to link the two experimental approaches.   


\section*{Discussion}
In this paper we presented a combined numerical and experimental study
exploring melting processes in binary lipid mixtures. In detail, we focused
on the melting of the single lipid components  \textit{DMPC-d54} and
\textit{DSPC} independently. Monte Carlo simulations linked FTIR and DSC
measurements to each other and therewith MC simulations proved another
time to be a useful and powerful tool in exploring melting transitions in
lipid membranes.

\

\textbf{FTIR, DSC and MC Simulations}

\

FTIR allowed us to measure the absorption spectra of lipid suspensions
consisting of MLVs with varying \textit{DMPC-d54} and \textit{DSPC} ratios in
dependence of temperature. In particular we probed the $CH_2$ and $CD_2$
vibrational modes of the lipid fatty acid chains. The deuterium is heavier
than the hydrogen atom which is why the symmetric and asymmetric bands of the
$CH_2$ and $CD_2$ stretch vibrations display absorptions at different
wavenumbers. This made it possible to distinguish the \textit{DMPC-d54} and \textit{DSPC} lipids. 

We applied two different methods to analyze the absorption spectra
measured. One was based on monitoring the temperature dependence of absorption
maxima corresponding to the $CH_2$ and $CD_2$ stretch vibrations,
respectively. The other one was built on following changes in the peak area of
difference spectra (\textit{Difference Spectra method}). We compared the corresponding experimental results to the
temperature dependent ratio of disordered chains of the single lipid species
using MC simulations of a simple model (see fig.~\ref{fig:fig3}). 

These numerical simulations were based on a Doniach model of the lipid
chain melting process assuming one ordered and one disordered lipid chain
state. Nearest neighbor interactions only were considered and the lipid chains
were arranged on a triangular lattice. Even though it is a minimalistic model which  neglects that lipid chains
may reflect different degrees of disorder and the non-consideration of a loss
of lattice order during the melting process the calculated fraction of
disordered lipid chains of the single lipid species describes the experimental
data obtained with the \textit{Difference Spectra method} well. 

Previously this model has been shown to be suited to describe heat
capacity profiles as measured with DSC \cite{sugar:TwoCompLipid,
hac:dmdsdiff, seeger:fluct05}. Ten unique parameters are needed which
were obtained from experimental heat capacity profiles. Six of them follow
directly from the heat capacity curves of the one component systems. These are
twice entropy and enthalpy changes and the corresponding cooperativity
parameters. The missing four parameters have to be found by fitting the heat
capacity profiles of the binary mixtures. They must be and they are the same
for all of the possible lipid ratios. Due to the errors in the $c_p$-profile
measurements the parameter determination is subject to deviations from the
ideal set of parameters. In our case we have taken the remaining four parameters
from the \textit{DMPC:DSPC} system \cite{hac:dmdsdiff}. This proved to be
valid, but small deviations due to this assumption cannot be excluded, too. 

Heat capacity profiles contain  macroscopic
information on the melting behavior such as changes in enthalpy and
entropy. However, DSC does not allow to determine the microscopic behavior of
the melting process. Additionally, it does not allow to investigate the
melting of the single components. It is only suited to measure 
the bulk melting behavior.  FTIR, however, provides a mean of studying the
melting of single lipid species and observes the microscopic details, but
lacks the possibility to probe macroscopic properties. The model
underlying the MC simulations
describes heat capacity profiles and predicts the findings from
FTIR well. Therefore, the MC simulations provide a mean to link the
information of the two experimental techniques.

\

\textbf{The Melting Process}

\

The melting process of single components is adequately described using the numerical
simulations. From the experimental and the simulated results and their
corresponding derivatives it becomes clear
that the lower temperature melting component (\textit{DMPC-d54}) melts also in
the binary mixture at a lower temperature than do the \textit{DSPC} lipids
(fig. ~\ref{fig:fig4}). Still, both lipid species
melt over the whole temperature regime and not only at distinct temperatures
at which fluctuations assigned to the single lipid components are enhanced. In general, the component which is
in the minority shows two maxima. At one of these temperatures the
strength of the fluctuations of the component which is in the majority is the
highest. This effect underlines the cooperativity of the melting process. The
melting of one lipid increases the probability of the melting of another lipid.

The temperatures at which the fluctuations of the two components is  strongest
as obtained from the simulations and the experiments, agree well (fig. ~\ref{fig:fig5}). Further, the
numerical simulations allowed to calculate the heat capacity values and the
disordered chain fractions simultaneously. Therefore, we were able to compare
the temperatures at which the heat capacity is in its maximum with the
temperatures determined from the derivatives of the disordered chain fractions. These
temperatures agree with each other. In fig.~\ref{fig:fig1} the heat capacity
profile of the equimolar mixture displays two maxima. The maximum at lower
temperature corresponds to the enhancement in fluctuations  of the \textit{DMPC-d54}
lipids and the one at higher temperatures to the one of the \textit{DSPC}
lipids. However, there are instances at which the heat capacity profile
displays only one maximum even though one can determine two temperatures using
the evolution of the single lipid disordered chain ratios. This is because DSC
describes the bulk behavior and single events might not be detectable since
they show only a small contribution to the overall melting. This needs
attention in discussing melting processes in biological membranes. In
the case of biological membranes it might even be difficult to distinguish the
melting processes from the baseline of the calorimetric scan. This means that
melting is present, but it cannot be detected at all using DSC. 

\ 

\textbf{FTIR: The Two Approaches}

\

In this paper we used two different techniques to analyze the FTIR data. These
were following the temperature dependence of the  absorption maxima of either
the symmetric $CH_2$ or the asymmetric $CD_2$ stretch vibrations and the
area change of one of the peaks of a difference spectra. 

Using the first method we found , in accordance to \cite{Leidy:latorgdomform},
that the absorption maximum of the $CH_2$ stretch vibrations of the
\textit{DSPC} lipids developed towards smaller wavenumbers when
 the \textit{DMPC-d54} started to display stronger fluctuations (see
 the right panels of 
fig.~\ref{fig:fig3}).  It has been
interpreted that in these cases the \textit{DSPC} lipids start to
order. Concluding from the data presented in this work and by Leidy et al. \cite{Leidy:latorgdomform} this ordering should be stronger than in
the presence of a complete \textit{solid ordered} membrane. The
wavenumber at which the absorption is at a maximum drops to wavenumbers even
smaller than at low temperatures. There is, however, no further supporting
evidence that lipids might be able to order stronger than at low temperatures.
This behavior is in disaccord of what we have found with the second method
and the Monte Carlo simulations. Therefore, it seems reasonable that the
Difference Spectra method yields results which describe the melting process
correctly and monitoring band shifts only can lead to incorrect results. 

\

\

\textbf{Biological Relevance}

\

The model of a biological membrane by Singer and Nicolson considers the biological membrane to be
homogeneous \cite{sing:fluidmos}. In the recent years this view has 
changed. A heterogeneous membrane model has overtaken this older
picture \cite{Jacobson:revfluidmosaic, Lagerholm:MicDom}. Different kinds of heterogeneities
can be present in biological membranes, such as protein aggregates or lipid
domains. Even if a membrane should be completely in the \textit{liquid
disordered} phase, lipids might not mix ideally and a heterogeneity
exists as seen in fig.~\ref{fig:fig1} and as already discussed elsewhere \cite{seeger:fluct05}. This statement is also true for membranes in the \textit{liquid
ordered} phase \cite{Fidorra:POPCceramcol}. The role of lipid membrane heterogeneities has been discussed
exhaustingly in the literature ~\cite{OpdenKamp:phospholipase, Lichtenberg:phospholipase,
Grainger:phospholipase, Cannon:RegCalciumChannel, Bolen:unstapkcactivation,
Dibble:laterheteroactpkc, Orr:PKC_phosphserin, Tang:lipstructCaATP,
Melo:domainconnectionreactions, Salinas:changesenzymeactivity,
Sackmann:triggprocmembrstruct}. Thereby, it has not only been pointed out that
the coexistence of solid ordered and liquid disordered phases might be
important \cite{OpdenKamp:phospholipase, Lichtenberg:phospholipase,
Grainger:phospholipase, Cannon:RegCalciumChannel, Bolen:unstapkcactivation}, but also the enrichment of single lipid species in certain
regions of the lipid membrane \cite{Leidy:DomainIndAct,
Dibble:laterheteroactpkc, Orr:PKC_phosphserin, Tang:lipstructCaATP}.

Melting of single components cannot be detected from the bulk melting in the
calorimeter. However, most of the publications cited here deduce phase
separation of natural membranes from calorimetric results. Therefore, details
of these processes might not be revealed. It is
also possible that the melting process in a biological membrane displays a
broad profile and the heat changes detected by a calorimeter cannot be
distinguished from the background signal.

In this context not only the various lipid
membrane phases have to be named, but also the possibility of enrichment of
certain lipid species in the certain lipid domains. Therefore, however, a
detailed understanding of lipid melting transitions and lateral structuring
need to be obtained.  The existence of heterogeneities depends on the
thermodynamical behavior of the different lipid species. As seen in this study
they melt over a broad temperature regime with particular temperatures at
which they display strong fluctuations in more complex lipid mixtures. These temperatures depend on the lipid
species, but also on its interactions with other kinds of lipid present. Melting processes might not be finished even though a heat
capacity profile might not indicate any such events. The sensitivity might be
not sufficient. FTIR spectroscopy, however, provides means to
distinguish the melting of single components.

A possible reason for the big variety in lipid species might be to ensure a
needed lipid membrane heterogeneity and therewith to play a role membrane function. This then provides an explanation why a
dependence of lipid synthesize in biological membranes on outer physical parameters such as pressure or
temperature is required.


\

\textbf{Acknowledgments}

\

M. Fidorra was funded by BioNet and the Villum Kann Rasmussen Foundation.


\end{document}